\documentclass[12pt]{article}

\usepackage{scicite}
\usepackage{times}

\usepackage{amsmath}
\usepackage{amsfonts}
\usepackage{amssymb}
\usepackage{graphicx}
\usepackage{float}
\usepackage{hyperref}
\usepackage{multirow}
\usepackage{rotating}

\usepackage{lineno}

\renewcommand{\baselinestretch}{1.15}

\newenvironment{sciabstract}{%
\begin{quote} \bf}
{\end{quote}}

\title{
Gender and the influence of research environment in topic selection of early-career faculty in STEM
}

\author{Lluís Danús$^{1,2}$, Robert H. Davis$^{3}$,\\
Roger Guimerà$^{4,1,\ast}$, Marta Sales-Pardo$^{1,\ast}$\\
\normalsize{$^1$Department of Chemical Engineering, Universitat Rovira i Virgili, Tarragona, Catalonia}\\
\normalsize{$^2$Annenberg School for Communication, University of Pennsylvania, Philadelphia, United States}\\
\normalsize{$^3$ Department of Chemical and Biological Engineering, University of Colorado,}\\ \normalsize{Boulder, Colorado, United States}\\
\normalsize{$^4$ICREA, Barcelona, Catalonia}\\
\normalsize{$^\ast$To whom correspondence should be addressed;}\\
\normalsize{ E-mail: roger.guimera@urv.cat, marta.sales@urv.cat} 
}

\date{}

\begin{document} 


\baselineskip18pt


\maketitle


\begin{sciabstract}
We study the influence that research environments have in shaping careers of early-career faculty in terms of their research portfolio. We find that departments exert an attractive force over early-career newcomer faculty, who after their incorporation increase their within-department collaborations, and work on topics closer to those of incumbent faculty. However, these collaborations are not gender blind: Newcomers collaborate less than expected with female senior incumbents. The analysis of departments grouped by fraction of female incumbents  reveals that female newcomers in departments with above the median fractions of female incumbents tend to select research topics farther from their department than female newcomers in the remaining departments ---a difference we do not observe for male newcomers. Our results suggest a relationship between the collaboration deficit with female incumbents and the selection of research topics of female early-faculty, thus highlighting the importance of studying research environments to fully understand gender differences in academia.
\end{sciabstract}


\section*{Introduction}

The topics of scientific research are constantly evolving\cite{kuhn14}: New topics gain attention while others languish in response to external pressures, including funding frameworks and societal needs
\cite{bromham16,wagner22}. For individual researchers, the selection of a portfolio of research topics is a critical decision that has direct impact in the evolution of their scientific careers. This is specially true for early-career scientists and young faculty, whose future professional stability hinges upon their early choices \cite{milojevic14}. Despite the importance of such early choices, we still know very little about which are the factors that affect the selection of research topics by individuals beyond global trends.

Changing and expanding the research portfolio is a common trait of scientific careers of researchers in STEM fields  \cite{battiston19}. However, when selecting possible research topics, individuals need to consider the trade-off between innovation and exploitation of exiting topics: While the former potentially has high rewards but implies high risk, the latter offers milder recognition but also involves lower risk \cite{aleta19}. 
Another factor that plays a role in shaping scientific careers is establishing collaborations: Collaboration with top-tier institutions and with prominent researchers increases impact and can help in career promotion \cite{jones08,sekara18}. However, because collaborative team sizes are largely driven by the amount of resources available, putting an emphasis on collaboration can lead to gender segregation in different fields of study within the same area \cite{duch12, duch16}. Indeed, gender is another factor that has a strong impact in the scientific career of individuals. Leaving leaky pipelines apart \cite{etzkowitz00}, female faculty often publish fewer articles and receive less grant money  \cite{duch12,lariviere13,boekhout21}, are promoted at a later career stage than their males colleagues \cite{boekhout21}, are more likely to experience issues when discussing authorship \cite{ni21}, and are given less credit for their contributions \cite{ross22}.

Unfortunately, studies looking at the evolution of scientific careers rarely consider the host institution or department as one of the factors playing an important role in the development of early-career faculty. Although formal collaborations, like coauthoring a publication, are the most recognizable way through which researchers share their ideas with each other, departments are collections of people who are exposed to similar scientific influences by, for example, attending the same seminars and informal meetings. Indeed, researchers within the same institution are more likely to collaborate \cite{jones08}. These collaborations are typically face-to-face, resulting in lower communication costs \cite{cummings05} and having a higher chance to spark creativity \cite{horvat22}. However, the effect that research environments have in shaping research careers and whether this effect has gender disparities have not been assessed.

\begin{figure}[t!]
\begin{center}
\includegraphics[width=0.7\textwidth]{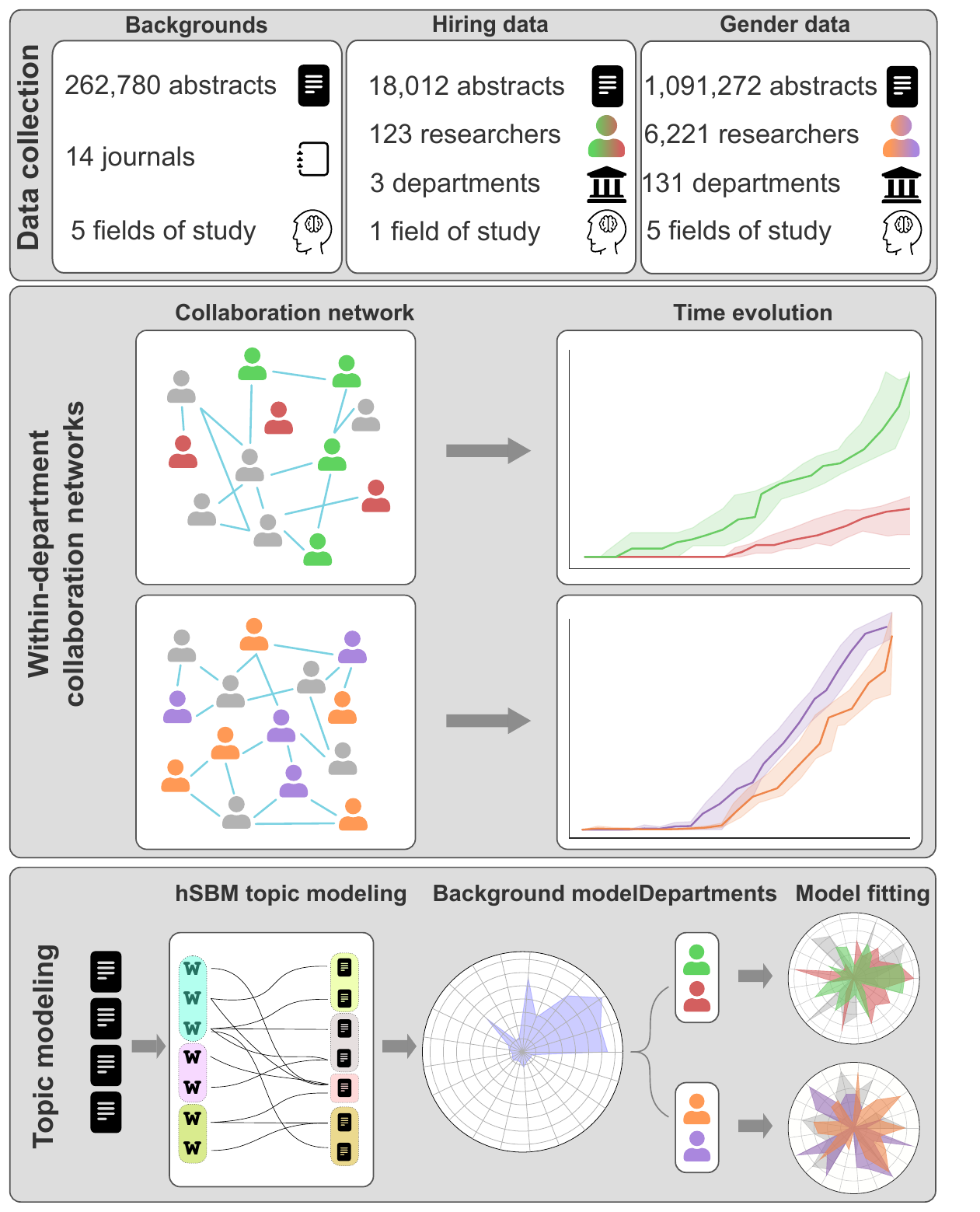}
\end{center}
%
\renewcommand{\baselinestretch}{1.0}
\caption{\footnotesize
{\bf Study design.}
We use three different types of datasets: (i)  ``background datatsets,'' which comprise all the abstracts of articles published in generalist journals in each of the fields of study (Materials and Methods); (ii) a ``hiring dataset'' that lists researcher who were offered a position at three institutions; and (iii) ``gender datasets'' for each field, which comprise the starting year and gender of all the faculty currently affiliated at selected highly ranked departments within the five fields of study (Materials and Methods) \cite{qs21}. 
First, for each of the researchers in the dataset, we look at the collaboration network  with incumbent department members before and after the offer/starting year. We then compare these collaborations between accepting and declining researchers (with offers from the same departments) and between male and female faculty over time. Second,  we build a bipartite network of documents and words for each background dataset and obtain research topics for each field. Using the list of publications of each researcher, we obtain the projection of each researcher onto topic space. This process allows us to make comparisons between researchers at different stages in their careers.}
\label{fig-abstract}
\end{figure}
Here, we aim precisely to cover this gap (Fig.~\ref{fig-abstract}). Our assumption is that departments and research institutions expose researchers to certain research questions and approaches, and they become incubators for novel ideas through collaboration among faculty members.  To investigate this, we consider two cohorts of early-career researchers.
The first cohort comprises young researchers who were offered an assistant professorship in a chemical engineering department; some of them declined and some accepted. The second cohort comprises early-career faculty in top departments in Europe and the United States in the fields of chemical engineering, biology, physics, mechanical engineering, and civil engineering. The analysis of the first cohort confirms that researchers who accepted an assistant professorship develop more collaborations with incumbent faculty in the department than those candidates who did not accept the offer, and it shows that the researchers who accepted offers are also more likely to converge towards the research topics of the department, even when discarding papers co-authored with other faculty in the department. The analysis of the second cohort reveals that, except in physics,  there is a systematic collaboration deficit between newcomer faculty (both male and female) and female senior incumbent faculty. We also find 
a striking difference between early-career male and female newcomer faculty across fields with the aforementioned collaboration deficit---female newcomer faculty in departments with below-median fractions of female incumbent faculty tend to converge less to department research topics than do female newcomers in departments with above-median fractions of female incumbent faculty. However, we see no such differences in the selection of research topics for male newcomers. 

Our findings show that, while departments exert an attractive force on early-career faculty, this force has different effects on male and female newcomer faculty. While more work still needs to be done to elucidate the reasons behind such disparities, our findings highlight the importance of studying  research environments to fully understand gender differences in academia.

\section*{Results}
\subsection*{Effects of joining a department on collaborations and research topics}

We start by quantifying the effect of joining a department on the collaborations and research topics of early-career faculty. To this end, we analyze the hiring history of three different chemical engineering departments in the United States between 2007 and 2017 (Fig.~\ref{fig-abstract}). Each of the three datasets contains a list of young researchers who were offered an assistant professor position in one of the departments and their response (accepted or declined), as well as the publications of those researchers and of all other faculty in the department, before and after the hiring offer. The candidates who accepted the offer give us the opportunity to analyze the effect of joining the department, in terms of both their number of within-department collaborators and their research topics. The researchers who declined allow us to control for potential confounding effects, such as field-wise shifts in research topics.

\begin{figure}[t!]
\includegraphics[width=\linewidth]{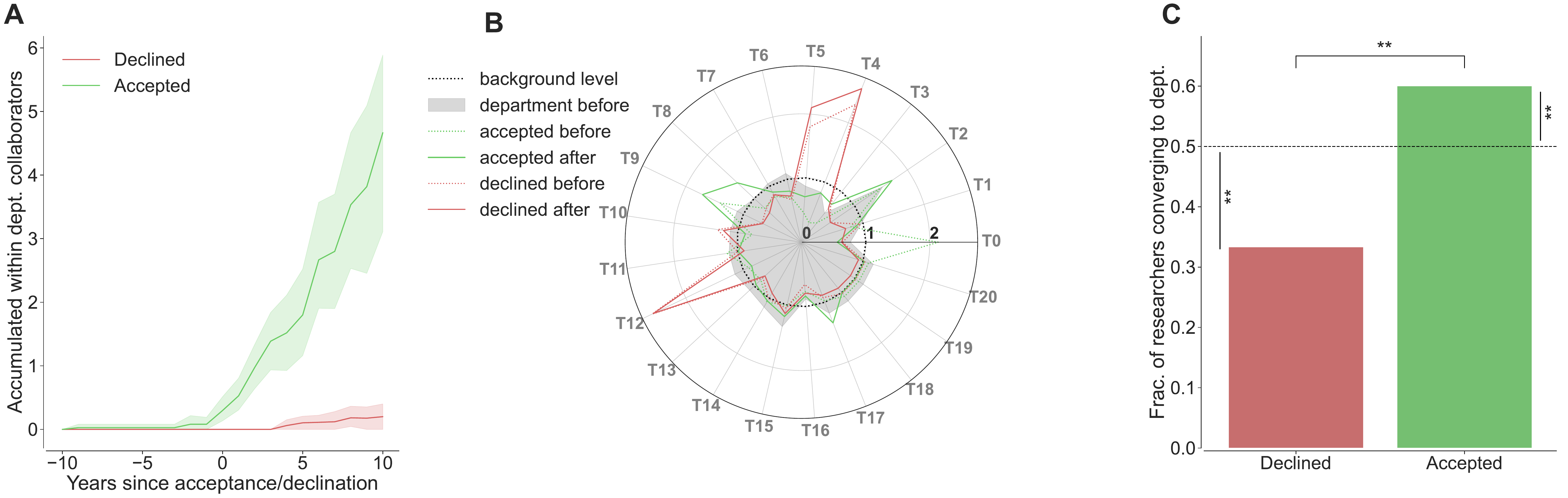}
\vspace{-7mm}
\renewcommand{\baselinestretch}{1.0}
\caption{\footnotesize
{\bf Changes in departmental collaborations and research topics after joining a department.}
{\bf (A)} Accumulated within-department distinct collaborators over time with respect to the year the offer to join the department was made for researchers who accepted (green) and who declined (red). The shadowed area corresponds to the 95\% confidence interval.
{\bf (B)} Comparison of topic distributions before (dotted lines) and after (solid lines) the year that the offer to join the department was made for researchers who accepted and for those who declined. The gray solid area corresponds to the background of the department at the time of the offer. All distributions are normalized with respect to the global distribution of the background dataset. The black dashed line represents the same topic proportion as that of the background.
{\bf (C)} Fraction of researchers whose topic distribution after the offer was made converges to that of the department -- researchers who declined the offer, red; researchers who accepted, green. 
Stars indicate statistical significance obtained from randomizing the accepted and declined labels in our dataset (***: 1\%, **: 5\%, *: 10\%, n.s.: not significant).
Note that in the analysis we exclude papers that are published in collaboration with department faculty, and, therefore, the effects we report are not a direct consequence of those collaborations.
}
\label{fig-hire-topics}
\end{figure}
As expected, the decision to join a department affects the number of collaborations of a researcher with members of that department. Figure~\ref{fig-hire-topics}A shows that, for the three departments considered here, a new faculty member joining the department increases the number of new collaborators within the department to an average of 5 collaborators after 10 years. By contrast, researchers who declined the offers barely collaborate with members of the department after declining.
We surmise that these new collaborations, as well as other forms of scientific socialization within the department, bring the early-career newcomer in contact with new methodologies, research questions, and ideas. To analyze the influence of the department on the research interests of the newcomers, we analyze the change in research topics  by means of topic modeling of article abstracts (see Methods), before and after the offer to join the department. In particular, we compare the distribution over topics of each researcher with that of the department (Fig.~\ref{fig-hire-topics}B). We observe that, before joining a department, the early-career faculty who accepted and those who declined were similarly close to the topic distribution of the department (Fig.~S1). However, after joining a department, newcomers tend to shift away from topics that are not popular in the department and towards others that are more prominent, even when we exclude direct collaborations (that is, direct co-authorship) with other department members. This result is in contrast to those who declined, who are less likely to converge towards the topics that are prominent in the department.

To quantify this effect, we compare the distribution of research topics of researchers and departments using information-theoretic metrics. In particular, we compute the Jensen-Shannon distance (see Methods) between the topic distributions of the early-career faculty (before and after the offer, accepted or declined) and that of the department (that is, of incumbent faculty before the offer). We exclude from this analysis all publications that involve co-authorship between the early-career faculty and incumbent faculty in the department; therefore, the convergence in topic distribution we observe is not an immediate effect of those collaborations.
Researchers converge towards the department if the change in distance  with respect to the department is negative (they become closer after acceptance or declination) and diverge from the department if the change is positive.
We find that the fraction of accepting early-career faculty who converge towards the department is 60\% (Fig.~\ref{fig-hire-topics}C), while the  fraction of declining early-career faculty that converge towards the department is only 32\%, both significant at the 5\% confidence level when compared to the null hypothesis of researchers randomly moving towards or away from the department and therefore converging or diverging with equal probability. The difference between the two fractions is also significant.

\subsection*{Departmental influence on male and female early-career researchers}

We have established that joining a department has a statistically significant effect on both the collaborations and the research topics of early-career faculty. However, Fig.~\ref{fig-hire-topics}C also reveals that 40\% of the newcomer faculty who joined one of the three departments considered in the first cohort did not converge towards it, suggesting that the influence of the department is not equal for everyone. Given that studies in the literature have shown stark differences in gender when it comes to scientific output, attribution and authorship \cite{duch12,lariviere13,ross22,ni21}, we further surmise that the research environment can also have a different effect depending on gender.


To test this hypothesis, we analyze data from the most prominent departments in North America and Europe (according to the QS World University Rankings \cite{qs21}) in five different fields: chemical engineering, physics, biology, mechanical engineering, and civil engineering (Fig.~\ref{fig-abstract}; Data; Supplementary Tables S1--5). For each one of these departments, we identified the tenure-track researchers (assistant, associate and full professors) and collected their publication and career data. With this information, we construct faculty-level collaboration networks and compute the topic distribution of each researcher. These data, together with information about gender and the year in which each researcher joined the department, allow us to compare the effects of the department on the newcomer faculty. Since we can expect more-established researchers to be less permeable to new research topics, even when joining a new department, we focus on researchers who had less than 30 papers published at the time of joining. Other than this restriction, we consider as newcomers all faculty who joined one of the departments considered after the year 2000, and as incumbents those faulty having joined before.


\begin{figure}[t!]
\centerline{\includegraphics[width=.75\linewidth]{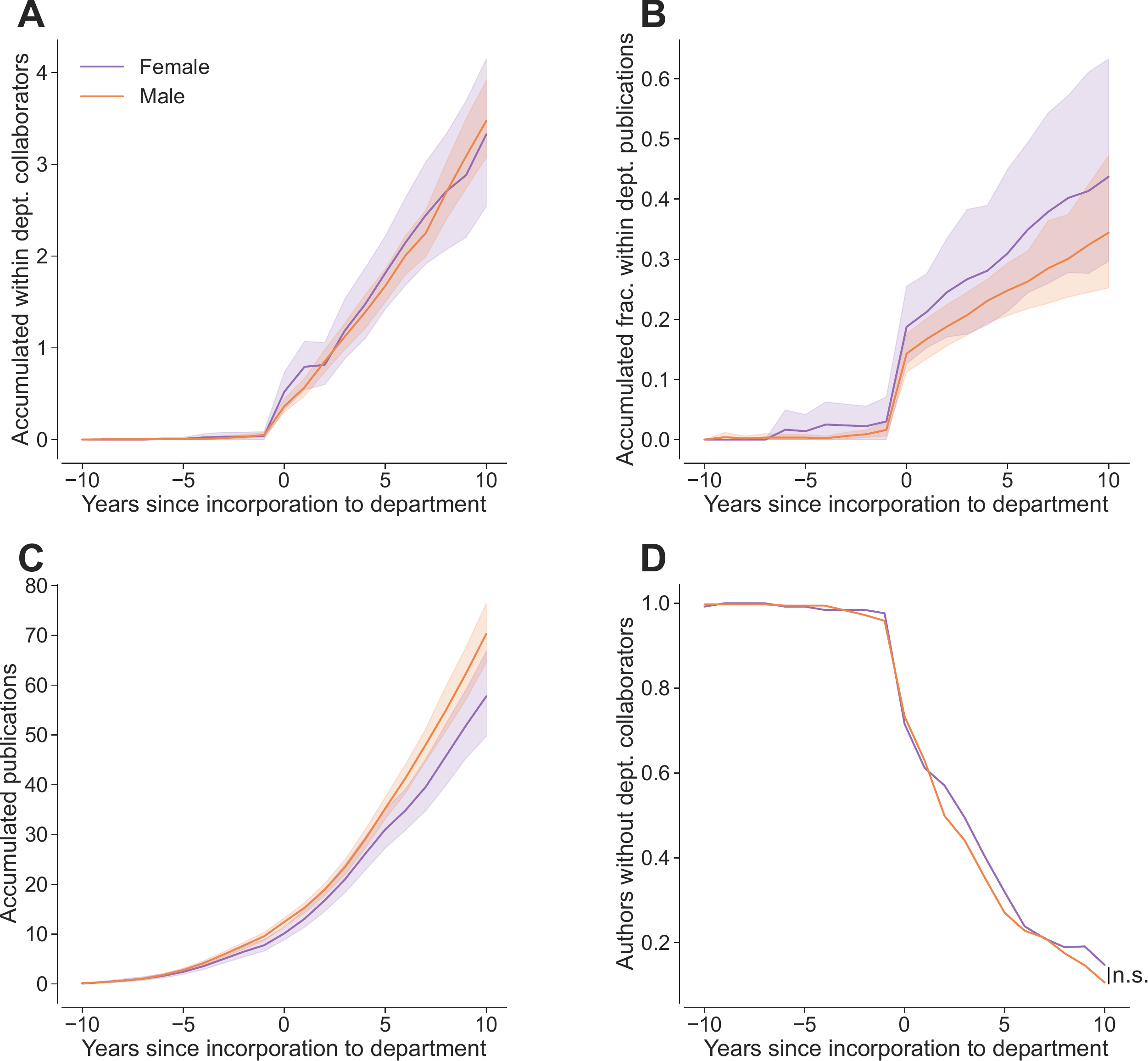}}
\renewcommand{\baselinestretch}{1.0}
\caption{\footnotesize
{\bf Overall absence of significant gender differences in within-department collaborations after joining a chemical engineering department.}
{\bf (A)} Accumulated within-department distinct collaborators with respect to the year of incorporation to the department for male (orange) and female (purple) researchers. The shadowed area corresponds to the 95\% confidence interval.
{\bf (B)} Fraction of publications that are with departmental collaborators.
{\bf (C)} Accumulated number of publications.
{\bf (D)} Fraction of authors with no collaborators within the department.
The same analyses for the fields on biology, physics, mechanical and civil engineering can be found in Supplementary Figs. S2-5.
}
\label{fig-gender-productivity}
\end{figure}
After joining a department, male and female researchers do not display significant differences across fields in the overall number of departmental collaborators or in the fraction of their publications that involve collaborations with other faculty in the department (Fig.~\ref{fig-gender-productivity}A-B for chemical engineering and Supplementary Figs.~S2-5 A-B for other fields). We also find that, while female newcomers tend to have a smaller output than male newcomers, differences are not significant in the majority of fields we study (except for biology). There are also no systematic differences in terms of the fraction of researchers who have no within-department collaborations (Fig.~\ref{fig-gender-productivity} C-D; Supplementary Figs. S2--5 C-D).


\subsubsection*{Gender differences in the selection of collaborations}

\begin{figure}[t!]
\centerline{\includegraphics[width=\linewidth]{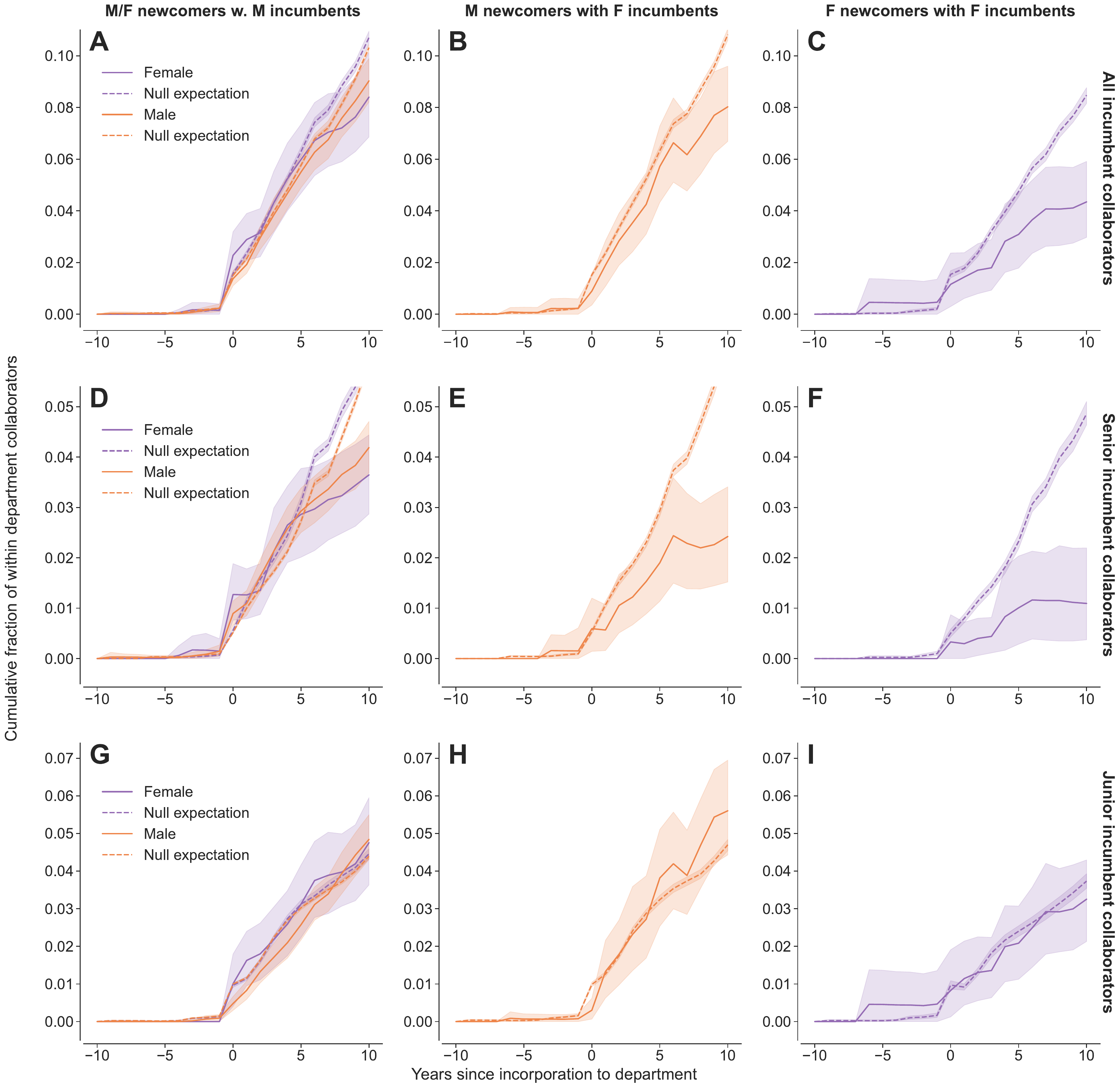}}
\renewcommand{\baselinestretch}{1.0}
\caption{\footnotesize
{\bf The gendered nature of within-department collaborations in chemical engineering departments.} Cumulative fraction of within-department collaborators by gender and career stage (purple represents females and orange represents males).  The shadowed area represents the 95\% confidence interval. The gray dashed line represents the expected fraction of  collaborations if we randomize female and male labels in each department.
{\bf (A)} Male and female newcomers with male incumbents.
{\bf (B)} Male newcomers with (junior and senior) female incumbents.
{\bf (C)} Female newcomers with (junior and senior) female incumbents.
{\bf (D)} Male and female newcomers with senior male incumbents.
{\bf (E)} Male newcomers with senior female incumbents.
{\bf (F)} Female newcomers with senior female incumbents.
{\bf (G)} Male and female newcomers with junior male incumbents.
{\bf (H)} Male newcomers with junior male incumbents.
{\bf (I)} Female newcomers with junior female incumbents.
For all other fields, see Supplementary Figs.~S6-S9.
}
\label{fig-collaborations-all}
\end{figure}
%
%
Despite the absence of differences described above for the overall number of collaborators of early-faculty within their new departments, we next show that there exist differences in the way male and female newcomers' collaborations are {\em distributed} among junior and senior, male and female incumbents. We focus on this distinction because senior incumbent faculty are in general more influential within departments, so that early-career faculty might favor collaboration with and be more affected by senior professors within the department. Since women are typically underrepresented in academic senior positions \cite{duch12}, we surmise that there might be effects mediated by the interaction of  gender and seniority.

%
%
Figure~\ref{fig-collaborations-all}A-C represents, for chemical engineering (see Supplementary Figs. S6-9A-C for other fields),  the cumulative fraction of within-department collaborators (that is, the number of collaborators of the early-career faculty members divided by the number of collaborators available), separated by the gender of the new faculty member and the gender of the incumbent collaborator, but ignoring the seniority of the incumbent. Female and male early-career faculty have indistinguishable fractions of male incumbent collaborators, and neither group deviates from the null expectation of selecting collaborators without taking gender into account (Fig.~\ref{fig-collaborations-all}A; Supplementary Figs. S6-9A). For chemical engineering, we observe that, as time goes by, both male and female newcomer researchers seem to collaborate with a lower fraction of female incumbent faculty than expected (Fig.~\ref{fig-collaborations-all}B-C). However, this observation is not consistent across fields (Supplementary Figs. S6-9B-C).

\begin{figure}[t!]
\centerline{\includegraphics[width=\linewidth]{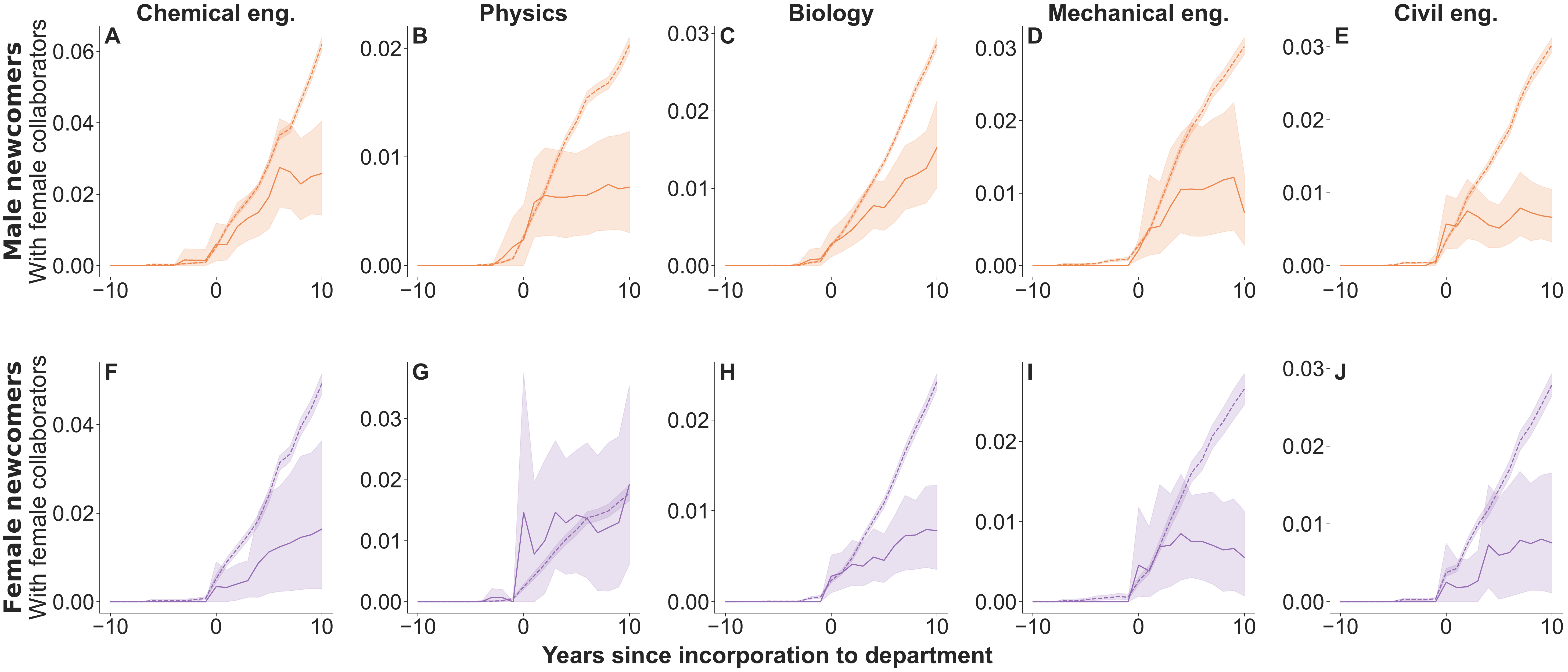}}
\renewcommand{\baselinestretch}{1.0}
\caption{\footnotesize
{\bf The collaboration deficit with senior female incumbents.} Cumulative fraction of within-department collaborators of senior female incumbent faculty with newcomer faculty separated by gender (purple: females; orange: males).  The shaded area represents the 95\% confidence interval. The dashed line represents the expected fraction of  collaborations if we randomize female and male labels in each department.
{\bf (A, F)} Chemical engineering.
{\bf (B, G)} Physics.
{\bf (C, H)} Biology.
{\bf (D, I)} Mechanical engineering.
{\bf (E, J)} Civil engineering.
}
\label{fig-collaborations-senior}
\end{figure}

Next, we introduce seniority into the analysis. In particular, we examine the effect of gender and seniority in collaborations by separating collaborations with senior incumbents (those who have a career at least 10 years longer than the early-career faculty member) and junior incumbents (those who have a career at most 10 years longer than the early-career faculty member). We find that, despite there being no systematic differences in overall collaboration rates between newcomers and incumbents disaggregated by gender, there are systematic gender  differences across fields in the way newcomers distribute collaborations between senior and junior incumbents (Fig.~\ref{fig-collaborations-all}D-I and Fig.~\ref{fig-collaborations-senior}).
Specifically, we find that as time goes by both male and female newcomers have a clear, systematic tendency to collaborate with a  lower fraction of senior female incumbents than expected  Fig.~\ref{fig-collaborations-senior}. There is only one exception to this rule---in physics departments, female newcomers collaborate with the expected fraction of female senior faculty. 
The reason why this deficit in collaboration with senior female incumbents is not always discernible in overall (senior and junior) collaboration rates with female incumbents is that, in some fields, newcomers collaborate with a slightly larger fraction of female junior incumbents than expected.  

By contrast, male and female newcomers collaborate with the expected fraction of male senior incumbents (except in biology in which all newcomers collaborate with a lower fraction than expected of senior researchers regardless of gender) (Fig.~\ref{fig-collaborations-all}D; Supplementary Figs. S6-9). We do not find any collaboration deficit with junior incumbents, either (Fig.~\ref{fig-collaborations-all}G-H; Supplementary Figs. S6-9G-H).


\subsubsection*{Gender and convergence to department topics}

So far, our results show that
there is a systematic gender bias in the way female and male newcomers establish collaborations with senior faculty within their departments (with the only exception of female newcomers in physics). 
%
We posit that the existence of such a gender bias could also imply that there is a gendered difference in the way departments exert influence over newcomer faculty. 
Indeed, senior faculty are expected to be the most influential among incumbent faculty, as they can offer more experienced counsel and guidance to young researchers. At the same time, incumbents of the same gender as the newcomers may exert more influence than incumbents of the opposite gender \cite{lockwood12, steele12, midgley21}.
%
%
In that case, we would expect that a deficit of collaborations with senior incumbent females may result in a loss of influence exerted by the department on newcomer females, a loss that would increase in departments with a larger fraction of incumbent females, where the deficit is felt more strongly.

\begin{figure}[t!]
\centerline{\includegraphics[width=\linewidth]{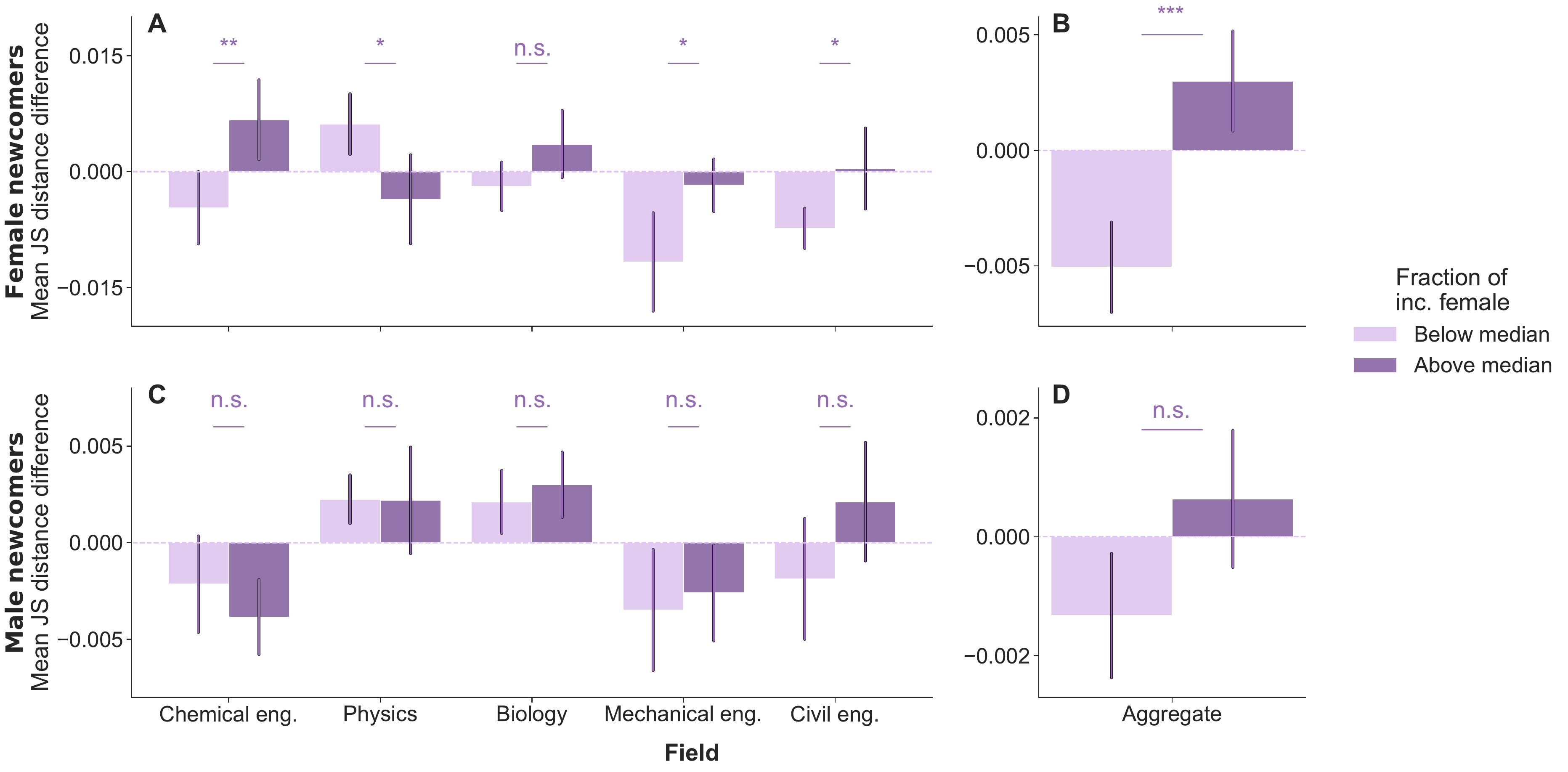}}
\renewcommand{\baselinestretch}{1.0}
\caption{\footnotesize
{\bf The fraction of incumbent females in a department is related to differences in the selection of research topics by female newcomers.} Mean Jensen-Shannon distance difference between newcomer and department research topics for newcomers before and 5 years after joining a department. For each field, we divide departments into two groups according to the fraction of female incumbent  faculty (above and below median, respectively; see legend). Each bar represents the mean over all departments in a given field and group (A and C), or in the aggregate for all fields (B and D). The upper row illustrates the results for female newcomers, while the bottom row presents data for male newcomers. Error bars represent the standard error. Stars indicate statistical significance of the  difference of means between the two groups obtained by randomizing the department membership to the above and below median groups
(***: 1\%, **: 5\%, *: 10\%, n.s.: not significant). Note that, in the analysis, we exclude papers that are in collaboration with department faculty, and, therefore, the effects we report are not a direct consequence of those collaborations (Methods). 
}
\label{fig-topics}
\end{figure}
To test this scenario in all fields, we separate departments into two groups accorting to whether they have above-median or below-median fractions of female incumbent faculty. Then, we look at the difference in distance between the research topics of newcomers and their department at the time of joining and five years after having joined the department. By separating newcomers by gender, we observe clear differences that are in agreement with our above scenario. In departments with an above-median fraction of incumbent females, newcomer females tend to converge less to the department topics than those in departments with a below-median fraction of incumbent female researchers (Fig.\ref{fig-topics}A). This indicates that in departments with higher fraction of females the influence of the department on female newcomers is weaker. This observation holds for all the fields we study except for physics; there, newcomer females joining departments with above-median fraction of female incumbent researchers become closer to the department. 
This result is also consistent with our scenario---since female newcomers in physics do not have a collaboration deficit with incumbent females (Fig.~\ref{fig-collaborations-senior}G), there is no loss in the influence of female incumbents on female newcomers therefore. Then, the higher the fraction of females, the more female newcomers would be expected to converge towards the topics of the department.

Note that, due to sample size, the statistical significance  for each field is moderate, but becomes highly significant if we pool all fields together (Fig.~\ref{fig-topics}B). By contrast, and also consistent with our scenario, we do not observe any significant differences across fields with the convergence towards the research topics of the department for male newcomer researchers, neither at an individual nor at an aggregate level (Fig.~\ref{fig-topics}C,D).

This difference between female and male newcomers is especially striking if we note that, at the time of joining the department, female newcomers are not further from the department in topic space than male newcomers are (Fig.~S10). Our analysis shows that, while we can observe little overall differences in the way within-department collaborations evolve for female and male newcomers, the nuance of how these collaborations are distributed affect how their research topics evolve differently after joining.

\section*{Discussion}
 
Our results show that working environments, and in particular hiring departments, exert a subtle yet quantifiable attractive force in the way young faculty shape their portfolio of scientific topics.
While constant contact with department colleagues can lead to collaborations that affect the topic-selection process, the convergence towards department topics is apparent even when we do not take direct collaborations into account. This finding unmistakably points to the fact that the local research environment permeates into the shaping of academic careers beyond strict paper co-authorship. This finding is in some ways surprising, because departments typically do not actively seek to affect researchers portfolios (quite on the contrary, top departments would likely expect their researchers to pursue independent, fruitful research careers), but they inevitably do have an impact on topic choices. 

Our results also show that female and male young faculty do not respond to departmental pressures in the same way. While women seem to have the same overall collaboration patterns as men, the deficit in collaborations with senior female faculty does not seem to have the same impact on both---in departments with a larger fraction of incumbent females, female newcomers (in fields with a deficit in collaborations with senior female incumbents) converge less to department topics, whereas the convergence of newcomer males in the same situation is independent of the fraction of incumbent women.  These rather stark differences are  both unexpected and  worrisome, and they could be explained by the fact that the influence of senior incumbent on newcomer faculty has a gender homophily dimension that only has discernible effects on women. Most broadly, our results point toward yet another gender difference in academia with unknown causes and unknown consequences down the road. 

In recent years, we have witnessed structural changes in many departments \cite{duch12,boekhout21}, with an increase in the fraction of female faculty, and yet large gender differences still exist in terms of promotion and credit---female researchers are typically given less credit for their contributions to research \cite{ross22}, and female faculty typically get promoted at later stages in their careers \cite{boekhout21}. These observations could suggest that women might look for different topic portfolios and different collaboration strategies as mechanisms to receive credit that could lead to future promotion, a mechanism that male faculty might feel less pressured to use. 

Studies from female academics in the 1990s already discussed that young women choosing the academic track did not necessarily find more support in departments with a larger presence of women faculty, and that some women felt the need to stand out on their own rather than clustering with other female faculty \cite{etzkowitz94,etzkowitz00}. It is unclear if female faculty currently feel the same way as in the 1990's and whether or not there are idiosyncratic differences that could explain the deficits in collaboration we observe. The obvious questions moving forward are therefore what are the reasons for these collaboration deficits  and what are the consequences of the differences in research topic selection that we observe? As many hiring departments make efforts and put policies in place to help reduce the gender gap in STEM departments, we cannot forget that hiring young female academics is not the endpoint but the start to close the gender gap in academia. The department is the research environment where young faculty should flourish, but right now this environment is not gender blind.

\section*{Materials and methods}

\subsection*{Data collection}

\paragraph{Corpus of articles used to obtain the background of research topics in chemical engineering} We collected the abstracts of all the 36,093 articles published in the \emph{AIChE Journal (AIChE)} and the \emph{Chemical Engineering Journal (CHEJ)} since their inception until 2021, available on Scopus. We selected these two journals because they are the most representative, broad-scope chemical engineering journals of North America and Europe, respectively.
\paragraph{Corpus of articles used to obtain the background of research topics in physics} We collected the abstracts of all the 120,990 articles published in \emph{Physical Review Letters (PRL)} since its inception until 2021, available on Scopus.

\paragraph{Corpus of articles used to obtain the background of research topics in biology} We collected the abstracts of all the 19,006 articles published in  \emph{Cell} since its inception until 2021, available on Scopus.

\paragraph{Corpus of articles used to obtain the background of research topics in mechanical engineering} We collected the abstracts of all the 45,210 articles published in the \emph{International Journal of Mechanical Sciences (IJMS)}, \emph{International Journal of Machine Tools and Manufacture (IJMTM)}, \emph{Journal of Sound and Vibration (JSV)}, \emph{International Journal of Plasticity (IJP)}, and in the \emph{Mechanical Systems and Signal Processing (MSSP)} since their inception until 2021, available on Scopus.

\paragraph{Corpus of articles used to obtain the background of research topics in civil engineering} We collected the abstracts of all the 41,481 articles published in the \emph{Building and Environment (BE)}, \emph{Coastal engineering (CE)}, \emph{Energy and Buildings (EB)}, \emph{Engineering Structures (ES)}, and the \emph{Journal of Construction Engineering and Management (JCM)} since their inception until 2021, available on Scopus.

\paragraph{Hiring data} Data about faculty hires were collected by R.H.D. and comprise the offers made to junior candidates by three different chemical engineering faculties in the United States. These data comprise more than 120 offers to researchers suitable to cover the position and the response to that offer. For each one of the candidates, we collected their full list of publications and retrieved the abstracts of these publications from Scopus. We also obtained faulty rosters of the three hiring departments at the moment the offer was made and retrieved full lists of publications and abstracts for these researchers from Scopus.

\paragraph{Faculty gender data} We first collected the list of the top 50 institutions across all fields from \emph{QS World University Rankings 2021} \cite{qs21} and selected those 136 departments based either in North America or Europe (Supplementary Tables S1-S5) for which an academic staff webpage was available. Then, for a total of 6,221 faculty in these departments, we downloaded their publication data, including abstracts and assigned them a binary gender based on their name.


\subsection*{Construction of the space of research topics}

\paragraph{Stopword removal} As a first step, we  pre-processed the abstracts to remove stopwords, those words such as prepositions or articles whose removal does not affect the meaning of a text. Instead of using a bag-of-words methodology, we followed the information-theoretical approach developed by \cite{gerlach19}, which allows us to identify stopwords in a context-dependent manner. Specifically, this methodology computes how informative a word is in the whole corpus by comparing the mutual information of the observed distribution of the occurrences of that word within the corpus to that of a random distribution of word occurrences. This approach allows us to define a threshold beyond which words are not considered informative. To better adjust this threshold, we ran iteratively our topic modeling under different values for informative words until groups composed by a majority of general words such as \emph{was, been, then...} vanished and all the words under this threshold for information content were removed, leading to a final corpus of 54 words per document on average. 

\paragraph{hSBM topic modeling}
With the background corpus of abstracts, we constructed a bipartite network with two types of nodes: words and documents. To obtain groups of documents and of words (research topics), we fit the  network to a Stochastic Block Model (SBM). SBMs are a class of  generative models for networks \cite{white76,nowicki01,guimera09}. In SBMs, nodes are assumed to belong to groups, and node-to-node connectivities are defined by the group memberships of the nodes alone. In particular, if nodes $i$ and $j$ belong to groups $g_i$ and $g_j$, respectively, then the probability that they are connected is given by a fixed quantity $p_{g_i g_j}$, which is identical for all other pairs in $g_i$ and $g_j$. The degree-corrected stochastic block model (SBM-DC) \cite{karrer11} is a variant of the SBM that allows for each node to have a different propensity to create links with other nodes, thus allowing nodes in the same group to have broad degree (connectivity) distributions despite having the same connectivity patterns.

Because group memberships are typically unknown, it is necessary to infer them from the observed connections in a given network. The most plausible partition of the nodes into groups (the partition that maximizes the Bayesian posterior over partitions  \cite{guimera09}) is also the one with the minimum description length (MDL), that is, the one that most compresses the observed connections \cite{peixoto14}. To obtain the MDL partition, one needs to specify a prior distribution over partitions. A hierarchical prior, which assumes that there is a nested hierarchy of groups of nodes, tends to yield shorter descriptions lengths than more uninformative priors. The SBM with such hierarchical priors is often referred to as the hierarchical SBM (hSBM) \cite{peixoto14}.

Approaches to topic modeling using SBMs overcome traditional topic modeling techniques such as LDA in recovering the text structure, as shown in \cite{gerlach18}. To compute the hSBM of the word-document network, we used the implementation in \href{https://graph-tool.skewed.de/}{graph-tool Python module}.

\paragraph{Background research topics} From the fit of the bipartite network of words and documents (abstracts) obtained from the background corpus to an hSBM we obtain a list of topics, that is a list of groups of words that are statistically equivalent in the way they are used in the documents. Each field required different levels in the hierarchy to obtain a good coarse-grained representation of the content of abstracts, resulting in an average of 20 topics per field.

\paragraph{Individual and departmental topic distributions} 
To build the topic distribution of an individual during a certain period of time, we consider all the pre-processed abstracts of articles authored by that researcher in that period of time. Because each word is associated to a topic, we can then construct a frequency vector of use of the $\mathcal{N}$ topics in the set of abstracts, which we call the topic distribution of an individual researcher.

To build a department's background topic distributions, we compute the distributions of senior faculty (i.e. those faculty that were at the department prior to 2000) according to the topics inferred by the hSBM. Once we have the topic distributions of all researchers, we calculate the average of all of them and use it as the average distribution before the beginning of the study. The general topic distribution for all the departments and fields under study is shown in (Fig.~S11-15).


\subsection*{Jensen-Shannon distance}

To calculate the distance between two topic distributions \emph{p} and \emph{q}, we compute the Jensen-Shannon distance, which is defined as
\begin{equation}
    {\rm JSD} = \sqrt{\frac{{\rm KL}(p||m) + {\rm KL}(q||m)}{2}} \;,
\end{equation}
where \emph{m} is the pointwise mean of \emph{p} and \emph{q} and ${\rm KL}$ is the Kullback-Leibler divergence between $p$ $({\rm or}\,q)$ and $m$. 
To prevent size effects in the estimate of the ${\rm KL}$ (an issue common to the estimation of information-theoretic metrics when the samples are small and some bins (topics) have very few observations), we compare, a random sample of 10 abstracts from each researcher to a random sample of the same size from a pool of papers by the authors of the department before the hiring process.


\bibliographystyle{Science.bst}

\section*{Acknowledgments}
This research was funded by project PID2022-142600NB-I00 from MCIN/ AEI/10.13039/501100011033 FEDER, UE and by the Government of Catalonia (2021SGR-633).

\subsection*{Authors contributions}
LD and RD obtained data. RD, RG and MSP designed research. LD, RG, and MSP analyzed data. All authors wrote the manuscript.

\subsection*{Competing interests}
The authors have no competing interests.

\subsubsection*{Data and materials availability}

Publication data are available from Scopus. Topic distributions of authors and departments needed to reproduce all results in the manuscript are available upon request.

\section*{Supplementary materials}

Figs. S1 to S15\\
Tables S1 and S5

\clearpage

\bibliography{ref-careers.bib}

\end{document}